\documentstyle[12pt,fleqn]{article} 
\begin{document}


\title{Statistical Coalescence Model \\
with Exact Charm Conservation}

\author{ M.I. Gorenstein$^{a,b}$, A.P. Kostyuk$^{a,b}$,
H. St\"ocker$^{a}$ and W. Greiner$^{a}$\\
$^a$ {\it Institut f\"ur Theoretische Physik, Universit\"at  Frankfurt,
Germany}\\
$^b$ {\it Bogolyubov Institute for Theoretical Physics,
Kyiv, Ukraine}
}



\maketitle

\begin{abstract}
The statistical coalescence model for the production of open and hidden
charm
is considered within the canonical ensemble formulation.
The data for the $J/\psi$ multiplicity in  Pb+Pb collisions at
158~A$\cdot$GeV are used for the model  prediction of the
open charm yield 
which has not yet been measured in these reactions. 
\end{abstract}


The charmonium states $J/\psi$ and $\psi^{\prime}$ have been measured in
nucleus-nucleus (A+A)
collisions at CERN SPS over the last 15 years by the NA38 and NA50
Collaborations.
This experimental program was motivated by a suggestion
 \cite{Satz1} to use the $J/\psi$ 
as a probe of the state of matter created in the early stage
of the collision. In this approach a significant suppression
of $J/\psi$ production relative to Drell--Yan lepton pairs
is predicted when going 
from peripheral to central Pb+Pb interactions at 158~A$\cdot$GeV.
This is originally
attributed to the formation of a quark-gluon plasma,
but could be also explained in microscopic hadron models as secondary
collision effects (see \cite{Sp} and references therein).

The statistical approach, formulated  in
Ref.\cite{Ga1}, assumes that
$J/\psi$ mesons are created at hadronization according to the available
hadronic phase-space. In this model the $J/\psi$ yield is 
{\it independent} of
the open charm yield. The model offers a natural explanation of the
proportionality of the $J/\psi$ and pion yields and the magnitude of the
$J/\psi$ multiplicity in hadronic and nuclear collisions.

Recently  the statistical coalescence model
\cite{Br1} and the microscopical coalescence model \cite{Le:00} were
introduced for the charmonium production.
Similar to the statistical model \cite{Ga1}, 
the charmonium states are assumed 
to be formed at the hadronization stage. 
However, they are produced as a coalescence of created earlier
$c$-$\overline{c}$ quarks and therefore 
the multiplicities of open and hidden charm
hadrons are {\it connected} in the coalescence models.
In Ref.~\cite{Br1} the
charm quark-antiquark pairs are assumed to be created at the early
stage of A+A collision and the average number of $c$-$\overline{c}$ pairs,
$N^{dir}_{c\overline{c}}$, 
is fixed by the model consideration based on the hard scattering 
approach.
The estimated number $N^{dir}_{c\overline{c}}$ seems to be larger than
the equilibrium hadron gas (HG) result.
This requires the introduction of a new parameter in the HG approach
\cite{Br1} --
the charm enhancement factor $\gamma_c$ 
(it was denoted as $g_c$ in Ref.\cite{Br1}). 
This is analogous to the
introduction of strangeness suppression factor $\gamma_s$ \cite{Raf1}
in the HG model, where the total strangeness observed is smaller than its
thermal equilibrium
value. Within this approach the open charm hadron yield is enhanced
by a factor $\gamma_c$ and 
charmonium yield by a factor $\gamma_c^2$ in comparison
with the equilibrium HG predictions.

The thermal HG calculations in Ref.\cite{Br1} are done 
within
the grand canonical ensemble (g.c.e.) formulation although 
the validity of the g.c.e. results for the charm hadron yield
was questioned by the authors of \cite{Br1} themselves.
As the total number of charm hadrons is expected to be smaller than
unity even for the most central Pb+Pb collisions an exact
charm conservation within the canonical ensemble (c.e.) should be
imposed\footnote{This was first suggested by K. Redlich and B. M\"uller
(quoted in Ref.\cite{Br1} and L. McLerran,
private communication).}. 
Note also that the c.e. formulation was successfully used in
Ref.\cite{BecD} to
calculate the open charm hadron abundances in $e^+e^-$ collisions
at $\sqrt{s}=91.2$~GeV with an experimental input  of the total open charm
production. 
In this letter stimulated by the above proposals
we consider the
c.e. HG formulation for the 
physical picture suggested in Ref.\cite{Br1}. The experimental value
of the $J/\psi$ multiplicity $\langle J/\psi\rangle $ will be then used
to predict the open charm yield within the statistical coalescence model.

\vspace{0.3cm}
The main assumption of Ref.\cite{Br1} is formulated as
\begin{equation}\label{Ncc}
N^{dir}_{c\overline{c}}~=~\frac{1}{2}~\gamma_c~N_O
+~\gamma_c^2~N_{H}~,
\end{equation}
where 
$N_O$ is the total thermal 
multiplicity of all open charm and anticharm mesons and (anti)baryons
and
$N_{H}$
is the total
thermal multiplicity of particles with hidden charm.
Note that open charm resonance states (not included in
\cite{Br1}) give essential contribution\footnote{We are thankful
to Braun-Munzinger and Stachel for pointing this out.} to
$N_O$. 
The number of directly produced
$c\overline{c}$ pairs $N^{dir}_{c\overline{c}}$ in the hard collisions 
is estimated in Ref.\cite{Br1} to be equal to
$N^{dir}_{c\overline{c}}\cong 0.17$ for Pb+Pb SPS collisions with
$N_p=400$ participants. 
This number is however not quite confident.
The average number of $c\overline{c}$ pairs created in nucleon--nucleon
collisions at $\sqrt{s}=17.3$~GeV was estimated from existing data as
$3\cdot 10^{-4}$
in Ref.\cite{GaD}. With a linear on $N_p$ extrapolation to  central
A+A collisions one obtains $N^{dir}_{c\overline{c}}\cong
0.06$ for $N_p=400$, but assuming that open charm production 
in central A+A collisions scales as 
$N_p^{4/3}$  an estimate  $N^{dir}_{c\overline{c}}\cong 0.35$ is
obtained for $N_p=400$  \cite{GaD}.
Note also that a
recent analysis of the dimuon spectrum measured in central Pb+Pb
collisions at 158 A$\cdot$GeV by NA50 Collaboration \cite{NA50}
suggests a significant enhancement of dilepton
production in the intermediate mass region (1.5$\div$2.5 GeV) over
the standard sources.
The primary interpretation attributes this observation to the
enhanced production of open charm \cite{NA50}:
about 3 times above the pQCD prediction
for the open charm yield in Pb+Pb collisions at SPS.

In Ref.\cite{Br1} $N_O$ and $N_{H}$ are calculated
in the g.c.e..
In the g.c.e. the thermal multiplicities of both open charm and
charmonium states are given as (Bose and Fermi effects are negligible):
\begin{equation}\label{gce}
N_j~=~ \frac{d_j
V~e^{\mu_j/T}}{2\pi^2}~T~m^2_j~K_2\left(\frac{m_j}{T}\right)~
\cong~d_j~V~e^{\mu_j/T}~\left(\frac{m_jT}{2\pi}\right)^{3/2}~\exp\left(-
\frac{m_j}{T}\right)~,
\end{equation}
where $V$ and $T$ correspond to the volume\footnote{To avoid further
complications we use ideal HG formulae and neglect excluded volume
corrections.} and
temperature
of HG system, $m_j$, $d_j$ denote particle masses and degeneracy
factors and $K_2$ is the modified Bessel function. The particle chemical
potential $\mu_j$ in Eq.(\ref{gce})
is defined as
\begin{equation}\label{mui}
\mu_j~=~b_j\mu_B~+~s_j\mu_S~+~c_j\mu_C~,
\end{equation}
where $b_j,s_j,c_j$ denote the baryonic number strangeness and
charm of particle $j$. The baryonic chemical potential $\mu_B$
regulates the baryonic density of the HG system whereas
strange $\mu_S$ and charm $\mu_C$ chemical potentials should be found
from the requirement of zero value for the total strangeness and charm
in the system (in our consideration we neglect small effects
of a non-zero electrical chemical potential). 

\vspace{0.3cm}
In the c.e. formulation (i.e. when the requirement of zero "charm charge" 
of the HG is used in the exact form) the thermal charmonium
multiplicities are still given by Eq.(\ref{gce}) as charmonium states
have zero 
charm charge.
The multiplicities (\ref{gce}) of open charm hadrons 
will however be multiplied by an additional 
'canonical suppression' factor (see e.g. \cite{Go1}).
This suppression factor is the same for all individual open charm states.
It leads to the total open charm multiplicity $N_O^{ce}$ in the c.e.: 
\begin{equation}\label{ce}
 N_O^{ce}~=~ N_O~\frac{I_1(N_O)}{I_0(N_O)}~,
\end{equation}
where
$N_O$ is the total g.c.e.
multiplicity of all open charm and anticharm mesons and (anti)baryons 
calculated with Eq.(\ref{gce})
and $I_0,I_1$ are the modified Bessel functions.   
For large open charm multiplicity $N_O>>1$ one finds
$I_1(N_O)/I_0(N_O)\rightarrow 1$ and therefore $N_O^{ce}\rightarrow
N_O$,
i.e. the g.c.e. and c.e. results coincide. For $N_O<<1$ one has
$I_1(N_O)/I_0(N_O)\cong N_O/2$ and
$N^{ce}_O\cong N_O \cdot N_O/2$,
therefore, $N_O^{ce}$ is strongly suppressed in comparison to
the g.c.e. result $N_O$ . 

Assuming the
presence of the charm enhancement factor $\gamma_c$ 
the statistical coalescence model within the c.e. should
now be formulated as:
\begin{equation}\label{Ncc1}
N^{dir}_{c\overline{c}}~=~\frac{1}{2}~\gamma_c~N_O~\frac{I_1(\gamma_c N_O)}
{I_0(\gamma_cN_O)}~
+~\gamma_c^2~N_{H}~.
\end{equation}
Therefore, the baryonic number, strangeness and electric
charge of the HG system are treated in our approach according to the
g.c.e. but charm is considered in the c.e.
formulation where the exact charge conservation is imposed.

The logic of Ref.\cite{Br1} is the following: 1) input
$N^{dir}_{c\overline{c}}$
number (it is assumed to be equal to 0.17 for $N_p=400$)
into Eq.(\ref{Ncc}); 2) calculate the $\gamma_c$ value; 3) obtain $J/\psi$
multiplicity as $\langle J/\psi \rangle = \gamma_c^2 N_{J/\psi}$,
where $N_{J/\psi}$ is given by Eq.(\ref{gce}).
Note that the second
term in both Eq.(\ref{Ncc}) and (\ref{Ncc1}) gives only a tiny
correction to the first term.
Therefore, $\gamma_c\cong 2N^{dir}_{c\overline{c}}/N_O$.

Our consideration differs from that of Ref.\cite{Br1} in three points.
First, we use  Eq.(\ref{Ncc1}) instead of (\ref{Ncc}).
Second, in our calculations we take into account all known particles
and resonances with   
open and hidden charm \cite{pdg}.
Third, we will proceed with Eq.(\ref{Ncc1}) in the
reverse way.
As the  $\langle J/\psi \rangle$ multiplicities
can be extracted from the NA50 data
on Pb+Pb collisions at 158 A$\cdot$GeV for
different values of $N_p$,
we start from the requirement:
\begin{equation}\label{Npsi}
\langle J/\psi \rangle~= ~\gamma_c^2~N_{J/\psi}^{tot}~,
\end{equation}
to fix the $\gamma_c$ factor.
In Eq.(\ref{Npsi}) the total $J/\psi$ thermal multiplicity
is calculated as 
\begin{equation}\label{dec}
N_{J/\psi}^{tot}=N_{J/\psi}~+~Br(\psi^{\prime})N_{\psi^{\prime}}~+~
~Br(\chi_1)N_{\chi_1}~+~
Br(\chi_2)N_{\chi_2}~,
\end{equation}
where $N_{J/\psi}$, $N_{\psi^{\prime}}$, $N_{\chi_1}$, $N_{\chi_2}$
are given by Eq.(\ref{gce}) and $Br(\psi^{\prime})\cong 0.54$,
$Br(\chi_1)\cong 0.27$, $Br(\chi_2)\cong 0.14$ are the decay branching
ratios of the excited charmonium states into $J/\psi$.  
Eq.(\ref{Ncc1}) will be used then to calculate the value of
$N^{dir}_{c\overline{c}}$. This value will be considered as a prediction
of the statistical coalescence model:  the open charm yield has not yet
been measured
in  Pb+Pb collisions at SPS.

We use two different
sets of the chemical freeze-out parameters:

\vspace{0.2cm}
 {\bf A}:~~~~
 $T=168$~MeV, $\mu_B=  266$~MeV, $\gamma_s = 1$~~~~
Ref.\cite{Br2};

\vspace{0.2cm}
{\bf B}:~~~~
$T=175$~MeV, $\mu_B=240$~MeV, $\gamma_s=0.9$~~ Ref.\cite{Bec}.

\vspace{0.2cm}
\noindent
They both were fixed
by the  HG model fit
to the hadron yields data
in Pb+Pb collisions at 158 A$\cdot$GeV
 (the inclusion of open charm and charmonium
states does not modify the rest of the hadron yields).
For the fixed number of participants $N_p$ 
the volume $V$ is defined then from $N_p=Vn_B$,
where $n_B=n_B(T,\mu_B,\gamma_s)$ is the baryonic density calculated in
the g.c.e.. 
With two sets of the chemical freeze-out parameters {\bf A} and 
{\bf B} we 
find $N_{J/\psi}^{tot}$ and $N_O$ values using 
Eq.(\ref{gce}), calculate $\gamma_c$
factors
from Eq.(\ref{Npsi}) and then calculate $N^{dir}_{c\overline{c}}$
from Eq.(\ref{Ncc1}). 

In the above c.e. consideration with exact charm conservation the $\gamma_c$
parameter regulates the {\it average}
number $N_{c\overline{c}}$ of $c\overline{c}$-pairs in the HG.
Therefore, $N_c=N_{\overline{c}}$ is restricted exactly (the c.e.),
but the value of $N_c+N_{\overline{c}}$ is restricted on average
(the g.c.e). 
The above c.e. calculations
are based therefore on the thermal model distribution
for probabilities 
to observe $0,1,2,...$ of $c\overline{c}$-pairs in the equilibrium
HG. One needs then an additional parameter
$\gamma_c$ to adjust these thermal probabilities to the
required number of $N^{dir}_{c\overline{c}}$.
Another way 
is to restrict also the $N_c+N_{\overline{c}}$ numbers in the
c.e. calculations and use
non-thermal probabilities to create $k=1,2,...$
of $c\overline{c}$-pairs in hard collisions. 
For the fixed number of $c\overline{c}$-pairs equal
to $k$,   
the average multiplicity of hidden charm can be approximately calculated
in the following way. 
We keep in the c.e. HG partition function the leading
terms only with 0 and 1 hidden charm particles and neglect all
configurations with 2,3,..,$k$ charmonium particles. This corresponds
to the expansion in powers of the small parameter  $N_H/(N_O/2)^2<< 1$. It
gives:
\begin{equation}\label{nhce}
\langle N_H \rangle_k~ \approx~ k^2 \frac{N_H}{(N_O/2)^2} ~,
\end{equation}
where  multiplicities $N_O$ and $N_H$ are calculated in the
g.c.e. using Eq.(\ref{gce}).

Because of the assumed hard scattering origin of the $c\overline{c}$
production 
the Poisson distribution $P(k)=f^k\exp(-f)/k!$ looks quite natural
($k=0,1,2,...$ is the number of pairs created, $f=
N_{c\overline{c}}^{dir}$
is the average number of pairs). The calculations with these
'dynamical' probabilities contain no additional free parameter.
All 'dynamical' information needed for the c.e. calculation is now given
by the value of $f$ (parameter $\gamma_c$ does not appear).
With Eq.(\ref{nhce}) the result for $J/\psi$ yield is:
\begin{equation}\label{pois}
\langle J/\psi \rangle
~\approx~ f(f+1)~\frac{N_{J/\psi}^{tot}}{(N_O/2)^2}~,
\end{equation}
where 
$N_{J/\psi}^{tot}$  is given by Eq.(\ref{dec}).
Using the experimental values for $\langle J/\psi
\rangle$ one obtains from Eq.(\ref{pois}) the average number
of $c\overline{c}$-pairs $f= N_{c\overline{c}}^{dir}$.    
The results appear to be rather
close to those obtained with thermal probabilities.
The reason of this fact is that states with $k=0$ and $k=1$
dominate in both thermal and 'dynamical' probability distributions.
To illustrate this lets consider an extreme choice:
the HG
states with $N_c=N_{\overline{c}}=1$ appear with
probability
$f=N^{dir}_{c\overline{c}}$, the HG states with $N_c=N_{\overline{c}}=0$
appear with probability $1-f$, and states with more than one
$c\overline{c}$-pairs are neglected.     
Under these restrictions  
the $J/\psi$ multiplicity becomes equal to:
\begin{equation}\label{psi}   
\langle J/\psi \rangle ~=~f~\frac{N_{J/\psi}^{tot}}{(N_O/2)^2 + N_{H}}~.
\end{equation}
One sees that Eq.(\ref{psi}) is close to Eq.(\ref{pois})
if $f<<1$ and $ N_H << (N_O/2)^2 $.
 From Eq.(\ref{psi}) one finds:
\begin{equation}\label{f}
f~=~\frac {\langle J/\psi\rangle}{N_{J/\psi}^{tot}}~\left[(N_O/2)^2 +
N_{H}\right]~=~
\gamma_c^2~\left[(N_O/2)^2 +
N_{H}\right]~.
\end{equation}
This coincides with Eq.(\ref{Ncc1}) at small values of $\gamma_c N_O$.

\vspace{0.3cm}
We present now the model calculations for central 
Pb+Pb interactions at 158~A$\cdot$GeV ($N_p=100\div 400$).
Using the  estimates for  experimental $J/\psi$ multiplicities
and assuming that the
system volume $V$ scales 
linearly with $N_p$ (i.e. $N_p=Vn_B(T,\mu_B,\gamma_s))$ we, first,
calculate  the thermal 
$J/\psi$ multiplicity $N_{J/\psi}^{tot}$ (\ref{dec}) -- including
the feeding from the excited charmonium states. Then we use Eq.(\ref{Npsi})  
to find the parameter $\gamma_c$. Finally, we calculate the predicted
values of $N_{c\bar{c}}^{dir}$
from Eq.(\ref{Ncc1}).
The results are presented in Tables 1 and 2
where the sets of the chemical freeze-out parameters
{\bf A}  and {\bf B} are respectively used.


\vspace{0.3cm}
A reliable extraction of the $J/\psi$ yields from the
published data appears to be  
non-trivial\footnote{We are thankful to K. Redlich and M.~Ga\'zdzicki for
the useful comments.},
Ref.\cite{GaJ} suggests
an approximately linear increase of $\langle J/\psi \rangle$
with $N_p$. The results for $\langle J/\psi \rangle$ presented
in Ref.\cite{Ga1} were evaluated
from the data of the NA50 Collaboration \cite{NA50new} using the
procedure described in \cite{APP}.
These results  for $\langle J/\psi \rangle$ are used as input for the
statistical coalescence
model analysis in Tables 1 and 2.
Assuming 
that $N^{dir}_{c\overline{c}}$ scales as $N_p^{\alpha}$
we find from Tables 1 and 2
a value of
$\alpha = 1.6\div 1.7$. 
This value is larger than $\alpha \cong 4/3$ expected
in the hard-collision model.
Although the values of $N_{J/\psi}^{tot}$,
$N_O$ and $\gamma_c$ are rather sensitive to the temperature parameter,
the model predictions for $N_{c\bar{c}}^{dir}$ remain essentially unchanged.


\begin{table}[hbt]
\begin{center}
Table 1 \\
\vspace{0.3cm}
\begin{tabular}{|c|c|c|c|c|c|c|}
\hline 
\rule[-3mm]{0mm}{10mm}
  &$\langle J/\psi \rangle \cdot 10^4$ & $N_{J/\psi}^{tot}\cdot
10^4$
      &$N_{O}$  & $\gamma_c$ &
           \multicolumn{2}{c|}{ $N_{c\bar{c}}^{dir}$ }  \\
                                            \cline{6-7}
$N_{p}$ & NA50 data & Eq.(\ref{dec}) & &  & Thermal & Poisson \\
 & Compil. \cite{Ga1} & Set {\bf A} & Set {\bf A}
    &Eq.(\ref{Npsi})& Eq.(\ref{Ncc1})&  Eq.(\ref{pois}) \\
\hline
 100 & 2.2$\pm$02 & 0.56 & 0.26 & 2.0\ & 0.066 & 0.064   \\
 200 & 3.9$\pm$0.2 & 1.1 & 0.52 & 1.9\ & 0.21\ & 0.20\   \\
 300 & 6.4$\pm$0.6 & 1.7 & 0.79 & 2.0\ & 0.46\ & 0.41\   \\
 360 & 6.9$\pm$0.7 & 2.0 & 0.94 & 1.9 &  0.57\ & 0.51\    \\
\hline
\end{tabular}
\end{center}
\end{table}


\begin{table}[htb]
\begin{center}
Table 2 \\
\vspace{0.3cm}
\begin{tabular}{|c|c|c|c|c|c|c|}
\hline 
\rule[-3mm]{0mm}{10mm}
  &$\langle J/\psi \rangle \cdot 10^4$ & $N_{J/\psi}^{tot}\cdot
10^4$
      &$N_{O}$  & $\gamma_c$ &
           \multicolumn{2}{c|}{ $N_{c\bar{c}}^{dir}$ }  \\
                                            \cline{6-7}
$N_{p}$ & NA50 data & Eq.(\ref{dec}) & &  & Thermal & Poisson \\
 & Compil. \cite{Ga1} & Set {\bf B} & Set {\bf B}
    &Eq.(\ref{Npsi})& Eq.(\ref{Ncc1})&  Eq.(\ref{pois}) \\
\hline
 100 & 2.2$\pm$02 & 1.1 & 0.39 & 1.4\ & 0.072 & 0.070   \\
 200 & 3.9$\pm$0.2 & 2.2 & 0.77 & 1.3\ & 0.23\ & 0.22\   \\
 300 & 6.4$\pm$0.6 & 3.3 & 1.17 & 1.4\ & 0.50\ & 0.45\   \\
 360 & 6.9$\pm$0.7 & 4.0 & 1.40 & 1.3 &  0.62\ & 0.55\    \\
\hline
\end{tabular}
\end{center} 
\end{table}

\vspace{0.3cm}
In conclusion, the statistical coalescence model with an exact charm 
conservation is formulated. The canonical ensemble suppression effects
are important for the thermal open charm yield even at $N_p=400$. 
These effects become
crucial when the number of participants $N_p$ decreases.
 From the $J/\psi$ multiplicity data in  Pb+Pb collisions at
158~A$\cdot$GeV the
open charm yield is predicted: $N_{c\bar{c}}^{dir} =0.5\div 0.6$
in central ($N_p=360$) collisions. 
An uncertainty of this prediction is mainly because
of uncertainties in different compilations of the $\langle J/\psi \rangle$
data.

It is interesting to compare our estimate 
$N_{c\bar{c}}^{dir}=0.5\div 0.6$ with 
results predicted in different model approaches.
The pQCD inspired models suggest the values of
$N_{c\bar{c}}^{dir} = 0.1\div 0.3$ 
(the value of $N_{c\bar{c}}^{dir}=0.17$ is
an estimate of Ref.\cite{Br1}). 
Much larger value of $N_{c\bar{c}} \cong 3.4$
is obtained in Ref.\cite{Le:00} within
the microscopic coalescence model. 
Even larger value of 
$N_{c\bar{c}} \approx 8$ is suggested in Ref.\cite{APP}
assuming the charm equilibration in the
quark-gluon plasma at the very early stage of Pb+Pb reaction. 

The statistical coalescence model
predicts also the $N_p$ dependence of $N_{c\bar{c}}^{dir}$
and the yields
of individual open charm states. All these predictions
of the statistical coalescence model
(the open charm yield has not been measured in Pb+Pb) can be tested
in the near future (measurements of the open charm are planned
at CERN).  This will require to specify more
accurately the  $\langle J/\psi \rangle$ data. 

The charm enhancement factor $\gamma_c$ found
from the  $\langle J/\psi \rangle$ data appears to be not much
different from unity and its value is rather sensitive to the temperature
parameter.
Therefore, both the statistical model of Ref.\cite{Ga1}
and the statistical coalescence model considered in the present
paper lead to
similar results for the $J/\psi$ yield. However,
the predictions of these
two models will differ greatly at RHIC energies:
according to \cite{Ga1} the $J/\psi$ to pion ratio is expected to be
approximately equal to its value at the SPS, but according to
the statistical coalescence model this ratio
should  increase very strongly. The predictions of the present model
for the RHIC energies will be presented elsewhere.

\vspace{0.3cm}
{\bf  Acknowledgments.}  The authors
are thankful to F. Becattini, P. Braun-Munzinger, K.A.~Bugaev, M. Ga\'zdzicki,
L. Gerland, L.~McLerran,
I.N.~Mishustin, G.C.~Nayak,  K.~Redlich and J.~Stachel for comments and
discussions.
We acknowledge the financial support of GSI and DAAD,
 Germany.
The research described in this publication was made possible in part by
Award \# UP1-2119 of the U.S. Civilian Research and Development
Foundation for the Independent States of the Former Soviet Union
(CRDF).

\end{document}